\input harvmac
\overfullrule=0pt
\input amssym

\def\Rop{{\Bbb R}}
\def\Cop{{\Bbb C}}
\def\vac{|0\rangle}
\def\n{\hbox{n}}
\def\A{{\cal A}}
\def\M{{\cal M}}
\def\L{{\cal L}}

\parindent=0pt
\parskip=6pt

\def\half{{\scriptstyle{1\over 2}}}

\Title{\vbox{\baselineskip12pt
}}
{\vbox{\centerline{From Dual Models to String Theory}
\bigskip
}}
\centerline{Peter Goddard}
\smallskip
\centerline{\it Institute for Advanced Study}
\centerline{\it Olden Lane, Princeton, NJ 08540, USA}
\vskip40truemm

{

\leftskip = 12 truemm
\rightskip = 12  truemm

A personal view is given of the  development of string theory out of dual models, including the analysis of the structure of the physical states and the proof of the No-Ghost Theorem, the quantization of the relativistic string, and the calculation of fermion-fermion scattering. 

}

\vskip 40truemm

\vfill\vfill
This article was written in response to the invitation of  Andrea Cappelli, Elena Castellani, Filippo Colomo and Paolo Di Vecchia to submit some personal reminiscences to the volume {\it The Birth of String Theory}.

\Date{}


\nref\Vena{G. Veneziano,
{\it Construction of a Crossing-Symmetric, Regge-Behaved Amplitude for Linearly Rising Trajectories},
Nuovo Cimento {\bf 57A} (1968) 190--197.}

\nref\Venb{G. Veneziano,
{\it Duality and Dual Models},
in {\it Proceedings of the 15th International Conference on High Energy Physics, Kiev, 26 August to 4 September,1970} (Kiev, 1972) 437--453.}

\nref\FGV{S. Fubini, D. Gordon and G. Veneziano,
{\it A General Treatment of Factorization in Dual Resonance Models},
Phys. Lett. {\bf 29B} (1969) 679--682.} 

\nref\Nama{Y. Nambu,
{\it Quark model and the factorization of the Veneziano amplitude}, 
in {\it Proceedings of the International Conference on Symmetries and Quark Models held at Wayne State University, June 18--20, 1969}, ed. R. Chand (Gordon and Breach, New York, 1970) 269--277, reprinted in {\it Broken Symmetry, Selected Papers of Y. Nambu}, ed. T. Eguchi and K. Nishijima (World Scientific, Singapore, 1995) 258--277.}

\nref\FVa{S. Fubini and G. Veneziano,
{\it Level structure of dual-resonance models,}
Nuovo Cimento {\bf 64A} (1969) 811--840.}

\nref\BM{K. Bardacki and S. Mandelstam,
{\it Analytic Solution of the Linear-Trajectory Bootstrap,}
Phys. Rev. {\bf 184} (1969)  1640--1644.}

\nref\Gl{F. Gliozzi,
{\it Ward-like Identities and Twisting Operator in Dual Resonance Models},
Lettere al Nuovo Cimento {\bf 2} (1968) 846--850.}

\nref\KN{Z. Koba and H.B. Nielsen,
{\it Reaction amplitude for n mesons: A Generalization of the Veneziano-Bardakci-Ruegg-Virasoro model,}
Nucl. Phys. {\bf B10} (1969) 633--655.}

\nref\Vira{M. Virasoro,
{\it Subsidiary Conditions and Ghosts in Dual-Resonance Models},
Phys. Rev. {\bf D1} (1970) 2933--2936.}

\nref\HBN{H.B. Nielsen,
{\it An Almost Physical Interpretation of the Integrand of the $n$-point Veneziano Model},
paper submitted to the 15th International Conference on High Energy Physics, 1970 and Nordita
preprint (1969).
}

\nref\Sa{L. Susskind,
{\it Harmonic-Oscillator Analogy for the Veneziano Model,}
Phys. Rev. Lett. {\bf 23} (1969) 545--547.}

\nref\Sb{L. Susskind,
{\it Structure of Hadrons Implied by Duality,}
Phys. Rev. {\bf D1} (1970) 1182--1186.}

\nref\Sc{L. Susskind,
{\it Dual-Symmetric Theory of Hadrons I,}
Nuovo Cimento {\bf 59A} (1970) 457--496.}

\nref\DVS{P. Di Vecchia and A. Schwimmer,
{\it The Beginning of String Theory:  a Historical Sketch} [{\tt arXiv:0708.3940}].}

\nref\Namb{Y. Nambu,
{\it Duality and Hadrodynamics},
(Notes prepared for the Copenhagen High Energy Symposium, unpublished, 1970), published in {\it Broken Symmetry, Selected Papers of Y. Nambu}, ed. T. Eguchi and K. Nishijima (World Scientific, Singapore, 1995) 280--301.}

\nref\Goto{T. Goto,
{\it Relativistic Quantum Mechanics of a One-Dimensional Continuum and Subsidiary Condition of Dual Resonance Model}, 
Prog. Th. Phys. {\bf 46} (1971) 1560--1569.}

\nref\FN{D.B. Fairlie and H.B. Nielsen,
{\it An Analogue Model for KSV Theory},
Nucl. Phys. {\bf B20} (1970) 637--651.}

\nref\Loa{C. Lovelace,
{\it $M$-Loop Generalized Veneziano Formula},
Phys. Lett. {\bf 32B} (1970) 703--708.}

\nref\Alessa{V. Alessandrini,
{\it A General Approach to Dual Multiloop Diagrams},
Nuovo Cimento {\bf 2A} (1970) 321--352.}

\nref\ELOP{R.J. Eden, P.V. Landshoff, D.I. Olive and J.C. Polkinghorne,
{\it The Analytic $S$-Matrix}, (Cambridge University Press, 1966).}

\nref\FGW{P.H. Frampton, P. Goddard and D.A. Wray,
{\it Perturbative Unitarity of Dual Loops}
Nuovo Cimento {\bf 3A} (1971) 755--762.}

\nref\Ga{P. Goddard, {\it Analytic Renormalisation of Dual One Loop Amplitudes},
Nuovo Cimento {\bf 4A} (1971) 349--362.}

\nref\Shapa{J.A. Shapiro,
{\it Renormalization of Dual Models,}
Phys. Rev. {\bf D11} (1975) 2937--2942.}

\nref\AAA{M. Ademollo, A. D'Adda, R. D'Auria, F. Gliozzi, E. Napolitano, S. Sciuto and P. Di Vecchia,
{\it Soft dilatons and scale renormalization in dual models},
Nucl. Phys. {\bf B94} (1975) 221--259.}

\nref\Lb{C. Lovelace,
{\it My Personal View,}
Contribution to this series of articles.}

\nref\Lc{C. Lovelace,
{\it Pomeron Form Factors and Dual Regge Cuts,}
Phys. Lett. {\bf 34B} (1971) 500--506.}

\nref\FVb{S. Fubini and G. Veneziano,
{\it Algebraic Treatment of Subsidiary Conditions in Dual Resonance Models}, 
Ann. Phys. {\bf 63} (1971) 12--27.}

\nref\DD{E. Del Giudice and P. Di Vecchia,
{\it Characterization of Physical States in of Dual-Resonance Models,}
Nuovo Cimento {\bf 70A} (1970) 579--591.}

\nref\BT{R. Brower and C.B. Thorn,
{\it Eliminating Spurious States from the Dual Resonance Model}, 
Nucl. Phys. {\bf B31} (1971) 163--182.}

\nref\Wak{M. Wakimoto, {\it Infinite-Dimensional Lie Algebras}, 
Translations of Mathematical Monographs {\bf 195} (American Mathematical Society, Providence, 2001).}

\nref\Thorna{C.B. Thorn, private communication.}

\nref\Thornb{C.B. Thorn, {\it A proof of the no-ghost theorem using the Kac determinant,} in {\it Vertex Operators in Mathematics and Physics} ed. J. Lepowsky, S. Mandelstam and I.M. Singer (Springer-Verlag, New York, 1985) 411-417.}

\nref\Kaca{V.G. Kac, {\it Contravariant form for the infinite-dimensional Lie algebras and superalgebras},
Lecture Notes in Physics {\bf 94} (1979) 441--445.}

\nref\FF{B.L. Feigen and D.B. Fuchs, {\it Skew-symmetric differential operators on the line and Verma modules over the Virasoro algebra,} Funct. Anal. and Appl. {\bf 17} (1982) 114.}

\nref\Virb{M. Virasoro,
{\it Alternative Constructions of Crossing-Symmetric Amplitudes with Regge Behavior},
Phys. Rev. {\bf 177} (1969) 2309--2311.}

\nref\Shapb{J.A. Shapiro,
{\it Electrostatic Analogue for the Virasoro Model,}
Phys. Lett. {\bf 33B} (1970) 361--.362.}

\nref\BGa{R.C. Brower and P. Goddard, 
{\it Generalised Virasoro Models},
Lettere al Nuovo Cimento {\bf 1} (1971) 1075--1081.}

\nref\Ram{P. Ramond,
{\it Dual Theory for Free Fermions},
Phys. Rev. {\bf D3} (1971) 2415--2418.}

\nref\NSa{A. Neveu and J.H. Schwarz,
{\it  Factorizable Dual Model of Pions,}
Nucl. Phys. {\bf 31} (1971) 86--112.}

\nref\NST{A. Neveu, J.H. Schwarz and C.B. Thorn,
{\it Reformulation of the Dual Pion Model,}
Phys. Lett. {\bf 35B} (1971) 529--533.}

 \nref\GW{P. Goddard and R.E. Waltz,
{\it One Loop Amplitudes in the Model of Neveu and Schwarz},
Nucl. Phys. {\bf B34} (1971) 99--108.}

\nref\DDF{E. Del Giudice, P. Di Vecchia and S. Fubini,
{\it General Properties of the Dual Resonance Model}, 
Ann. Phys. {\bf 70} (1972) 378--398.}

\nref\BGb{R.C. Brower and P. Goddard,
{\it Physical States in the Dual Resonance Model} in {\it Proceedings of the International School of Physics ``Enrico Fermi'' Course LIV} (Academic Press, New York and London, 1973) 98--110.}

\nref\BGc{R.C. Brower and P. Goddard,
{\it Collinear Algebra for the Dual Model},
Nucl. Phys. {\bf B40} (1972) 437--444.}

\nref\GT{P. Goddard and C.B. Thorn,
{\it Compatibility of the Dual Pomeron with Unitarity and the Absence of
Ghosts in the Dual Resonance Model},
Phys. Lett. {\bf 40B} (1972) 235--238.}

\nref\Bro{R.C. Brower,
{\it Spectrum-Generating Algebra and No-Ghost Theorem in the Dual Model},
Phys. Rev. {\bf D6} (1972) 1655--1662.}

\nref\Sch{J.H. Schwarz,
{\it Physical States and Pomeron Poles in the Dual Pion Model}, 
Nucl. Phys. {\bf B46} (1972) 61--74.}

\nref\BF{R.C. Brower and K.A. Friedman,
{\it Spectrum-Generating Algebra and No-Ghost Theorem for the Neveu-Schwarz Model},
Phys. Rev. {\bf D7} (1973) 535--539.}

\nref\Bor{R.E. Borcherds,
{\it Monstrous Moonshine and Monstrous Lie Superalgebras}
Invent. Math. {\bf 109} (1992) 405--444.}

\nref\Gicm{P. Goddard
{\it The Work of R.E. Borcherds},
in {\it Proceedings of the 	International Congress of Mathematicians, Berlin 1998.}
(Deutscher Mathematiker-Vereinigung, 1998) 99--108 [{\tt math/9808136}]}

\nref\GRT{P. Goddard, C. Rebbi and C.B. Thorn,
{\it Lorentz Covariance and the Physical States in the Dual Resonance Model,}
Nuovo Cimento {\bf 12A} (1972) 425--441.}

\nref\CM{L.N. Chang and F. Mansouri, 
{\it Dynamics Underlying Duality and Gauge Invariance in the Dual-Resonance Models},
Phys. Rev. {\bf D5} (1972) 2535--2542.}

\nref\MN{F. Mansouri and Y. Nambu,
{\it Gauge Conditions in Dual Resonance Models}, 
Phys. Lett. {\bf 39B} (1972) 375--378.}

\nref\GGRT{P. Goddard, J. Goldstone, C. Rebbi and C.B. Thorn,
{\it Quantum Dynamics of a Massless Relativistic String,}
Nucl. Phys. {\bf B56} (1973)109--135.}

\nref\Dirac{P.A.M. Dirac,
{\it Generalized Hamiltonian Dynamics},
Proc. Roy. Soc. {\bf A246} (1958) 326--332.}

\nref\Ma{S. Mandelstam,
{\it Interacting String Picture of Dual Resonance Models},
Nucl. Phys. {\bf 64} (1973) 205--235.}

\nref\JMS{J.-M. Souriau,
{\it Du bon usage des \'elastiques,}  
Journ. relat., Clermont-Ferrand, (1973).}

\nref\SW{R.F. Streater and A.S. Wightman,
{\it PCT, Spin and Statistics and All That}
(W. A. Benjamin, New York, 1964) p. 31.}

\nref\Thornc{C.B. Thorn,
{\it Embryonic Dual Model For Pions And Fermions},
Phys. Rev. {\bf D4} (1971) 1112--1116.}

\nref\Scha{J.H. Schwarz,
{\it Dual quark-gluon model of hadrons}, 
Phys. Lett. {\bf 37B} (1971) 315--319.}

\nref\CO{E.F. Corrigan, and D.I. Olive
{\it Fermion meson vertices in dual theories,}
Nuovo Cimento {\bf 11A} (1972) 749--773.}

\nref\CGa{E.F. Corrigan and P. Goddard,
{\it Gauge Conditions in the Dual Fermion Model,}
Nuovo Cimento {\bf 18A} (1973) 339--359.}

\nref\BO{L. Brink and D. Olive,
{\it Recalculation of the Unitary Single Planar Dual Loop in the Critical Dimension of Space Time,}
Nucl. Phys. {\bf B58} (1973) 237--253.}

\nref\OS{D. Olive and J. Scherk,
{\it Towards Satisfactory Scattering Amplitudes for Dual Fermions,}
Nucl. Phys. {\bf B64} (1973) 334--348.}

\nref\SWu{J.H. Schwarz and C.C. Wu,
{\it Evaluation Of Dual Fermion Amplitudes,}
Phys. Lett. {\bf 47B} (1973) 453--456.}

\nref\Grev{P. Goddard, 
{\it Dual Resonance Models},
Supplement au Journal de Physique, Tome {\bf 34}, Fasc. 11--12 (1973)
Cl 160--166.}

\nref\CGc{E.F. Corrigan and P. Goddard,
{\it Absence of Ghosts in the Dual Fermion Model},
Nucl. Phys. {\bf B68} (1974) 189--202.}

\nref\Tp{C.B. Thorn, private communication (April, 1973).}

\nref\SchRep{J.H. Schwarz, 
{\it Dual Resonance Theory,}
Phys. Rep. {\bf 8} (1973) 269--335.}

\nref\CGOS{E.F. Corrigan, P. Goddard, D.I. Olive and R.A. Smith,
{\it Evaluation of the Scattering Amplitude for Four Dual Fermions,}
Nucl. Phys. {\bf B67} (1973) 477--491.}

\nref\GSO{F. Gliozzi, J. Scherk and D. Olive,
{\it Supergravity and the Dual Spinor Model,}
Phys. Lett {\bf 65B} (1976) 282--286.}

\nref\Mb{S. Mandelstam,
{\it Interacting String Picture of the Neveu-Schwarz-Ramond Models},
Nucl. Phys. {\bf 69} (1974) 77--106.}

\nref\AP{A.M. Polyakov,
{\it Quantum Geometry of Bosonic Strings,}
Phys. Lett. {\bf B103} (1973) 207--210.}

\nref\RT{R. Giles and C.B. Thorn,
{\it Lattice Approach to String Theory},
Phys. Rev. {\bf D16} (1977) 366--386.}

\nref\Olive{D. Olive,
{\it Dual Models,}
in  {\it Proceedings of the 17th International Conference on High Energy Physics, London, July,1974} (Didcot, 1974) I 269--280.}

\nref\Gb{P. Goddard,
{\it The Connection between Supersymmetry and Ordinary Lie Symmetry 
      Groups,}
Nucl. Phys. {\bf B88} (1975) 429--441.}

\nref\GHP{P. Goddard, A.J. Hanson and G. Ponzano,
{\it The Quantization of a Massless Relativistic String in a Time-like Gauge,}
Nucl. Phys. {\bf B89} (1975) 76--92.}

\nref\GOa{P. Goddard and D.I. Olive,
{\it Algebras, Lattices and Strings}
in {\it Vertex Operators in Mathematics and Physics: Proceedings of 
a Conference November 10--17, 1983} (ed. J. Lepowsky, 
      S. Mandelstam and I. Singer) {\it Publications of the Mathematical Sciences 
      Research Institute, Berkeley, No. 3} (Springer-Verlag, 1984) 51--96.}
      
\nref\GNO{P. Goddard, J. Nuyts and D.I. Olive,
{\it Gauge Theories and Magnetic Charge,}
Nucl. Phys. {\bf B125} (1977) 1--28.}

\nref\AW{A. Kapustin and E. Witten,
{\it Electric-Magnetic Duality And The Geometric Langlands Program,}
{\tt hep-th/0604151.}}


\newsec{A Snapshot of the Dual Model at Two Years Old}

When particle physicists convened in Kiev at the end of August 1970, for the 15th International Conference on High Energy Physics, the study of dual models was barely two years old.
Gabriele Veneziano, reporting on the precocious subject   that he had in a sense created \refs{\Vena}, characterized it  as ``a very young theory still looking for a shape of its own and for the best direction in which to develop. On the other hand duality has grown up considerably since its original formulation. Today it is often seen as something accessible only to the `initiated'. Actually theoretical ideas in duality have evolved and changed very fast. At the same time, they have very little in common with other approaches to particle physics''  \refs{\Venb}. Nearly forty years later, his comments might still be thought by some to apply, at least in part.

In describing the theoretical developments, Veneziano stressed the progress that had been made in understanding the spectrum of the theory, obtained by factorizing the multiparticle generalizations of his four-point function in terms of the states generated from a vacuum state, $\vac$, by an infinite collection of harmonic oscillators, $a_n^\mu$, labeled by both an integral mode number, $n$, and a Lorentz index, $\mu$, \refs{\FGV, \Nama},
\eqn\acomm{
[a_m^\mu,a_n^\nu]=mg^{\mu\nu}\delta_{m,-n}; \qquad a^{\mu\dagger}_n=a^\mu_{-n};
\qquad a_n^\mu\vac=0,\quad\hbox{for } n>0.
}
Here $g^{\mu\nu}$ is the (flat) space-time metric, taken to have signature $(-, +,+,+)$ and, as Veneziano commented, the time-like modes, $a^0_n$, meant that theory potentially contained negative-norm or `ghost' states, which would lead to unphysical negative probabilities. [In this article I shall use the term `ghost'  to refer to a negative norm state rather than the fields associated with gauge invariance in functional approaches to quantization.]

A similar potentiality exists in the covariant approach to quantum electrodynamics, but the ghost states there are removed by a condition that follows from the electromagnetic  gauge invariance of the theory. A condition that removed some of the ghosts had been found \refs{\FVa,\BM} for the dual model, and this was associated with an SO(2,1) invariance \refs{\Gl} or, equivalently, M\"obius invariance in 
terms of the complex variables introduced by Koba and Nielsen \refs{\KN} to describe the $n$-point function in a symmetrical way.  But, this condition is not sufficient to remove all the ghosts, essentially because an infinity of such conditions is needed corresponding to the infinite number of time components, $a^0_n$. 

All was not lost, however. Virasoro had found \refs{\Vira}  that, when a parameter of the theory (the intercept of the leading Regge trajectory) $\alpha_0=1$, the M\"obius invariance enlarges to an infinite-dimensional symmetry corresponding to the conformal group in two dimensions, the complex plane of the Koba-Nielsen variables. For $\alpha_0=1$, the physical states of the theory satisfy a corresponding infinity of conditions, which stood a chance of eliminating all of the unwanted ghost states, but a proof of this was lacking. Further, the choice $\alpha_0=1$ brought with a problem: the masses, $M$, of the states in theory satisfy 
\eqn\spectrum{
\half M^2 =\sum_{n=1}^\infty\sum_{\mu=0}^3 N_n^\mu n -\alpha_0,}
for a state with occupation number $N_n^\mu$ in the $n$-th mode corresponding to the oscillator $a_n^\mu$,
and so the lowest state, with all occupation numbers zero, has $M^2=-2\alpha_0$. If $\alpha_0=1$, 
which we shall in general take to be the case  in what follows, so that Virasoro's conditions obtain, the lowest state has negative mass squared, {\it i.e.} it is  a tachyon: the cost of having a chance of eliminating ghosts was the presence of a tachyon!

In his report \refs{\Venb}, Veneziano also surveyed the attempts that had been made to interpret what had been learnt about the dual model and its spectrum in terms of a field theoretic, or some more directly physical, picture.  Holger Nielsen, in a paper \refs{\HBN} submitted to the conference and circulated the previous year, argued, first, that the integrand for the dual amplitude could in practice be calculated by using an analogue picture, formulated in terms of the heat generated in metallic surface into which currents, corresponding to the momenta of the external particles, are fed and, second,  that the dual amplitudes might be viewed as the approximate description of the contribution of a class of very complicated Feynman diagrams in a field theory, which he argued should be two-dimensional in some sense. Hadrons he described as `threads', the history of whose propagation was described by the two-dimensional surfaces. 

`String' interpretations had also been developed by Yoichiro Nambu \refs{\Nama} and Leonard Susskind
\refs{\Sa, \Sb, \Sc}. They each deduced that the dual model could be pictured in some sense as an oscillating string, because the spectrum   \spectrum\ of the dual model can be viewed as being  formed from energies associated with a basic oscillator and all its higher harmonics (or more precisely with one such oscillator for each dimension of space-time). At this stage, an intrinsically defined geometric action principle to govern the dynamics of the string had not been proposed. As time evolves, the string describes a surface (or `world sheet' \refs{\Sc}), $x^\mu(\sigma,\tau)$, where $x^\mu$ satisfies the wave equation as a function of $\sigma$ and $\tau$, seen as following from an action of a familiar quadratic form,
\eqn\quadratic{
\A_O\propto\int \L _O\,d\sigma d\tau,\qquad \L_O= {1\over 2}\left({\partial x\over\partial \sigma}\right)^2-
{1\over 2}\left({\partial x\over\partial \tau}\right)^2.}

As Di Vecchia and Schwimmer put it, ``using plausible arguments [Nambu, Nielsen and Susskind] obtained expressions similar to the N-point (tree) amplitudes'' \refs{\DVS} from this formalism, but, just before the Kiev Conference, elsewhere Nambu was considering whether it might not be more satisfactory to replace the quadratic action \quadratic\ with one that was intrinsically geometric, and he proposed the (Lorentzian) area,
\eqn\LNG{
\A_{NG} = -{T_0\over c} \int \L_{NG} \,d\sigma d\tau,
\qquad\L_{NG} = \sqrt{\left({\partial x\over\partial\sigma}\cdot
{\partial x\over\partial\tau}\right)^2 -
\left({\partial x\over\partial\sigma}\right)^2
\left({\partial x\over\partial\tau}\right)^2}.}
Being intrinsically geometric, this action is of course invariant under arbitrary changes of the parameters $\sigma$ and $\tau$. Nambu's initial discussion of this action  was in notes  \refs{\Namb} prepared for a symposium in Copenhagen in August 1970 that, in the event, he was unable to attend. [Although knowledge of this aspect of the content of the notes spread by word of mouth over time, the notes were not generally available until the publication of Nambu's {\it Selected Papers} in 1995.] Nambu's proposal of \LNG\  was made rather {\it en passant} and he did not discuss the consequences of the action in his notes beyond remarking that it obviously leads to nonlinear equations of motion. The action was discussed somewhat further by Goto \refs{\Goto} some months later, leading to $\A_{NG}$  being designated the Nambu-Goto action. 
We shall discuss the further development of the understanding of $\A_{NG}$ in Section 7.

Returning to Nielsen's analogue approach to calculating dual amplitudes, he had developed it  further with David Fairlie in a published paper \refs{\FN}, which showed how one-loop amplitudes could be obtained, as well as the tree amplitudes we have been discussing up to now. The analogue approach was also used by Lovelace \refs{\Loa} and by Alessandrini \refs{\Alessa} to motivate the mathematical constructions needed to calculate multiloop amplitudes. However, as Di Vecchia and Schwimmer 
\refs{\DVS} have emphasized, the initial development of the theory was based within the framework of $S$-matrix theory, rather than the quantization of the dynamics of a `string'. 

\newsec{Analyticity, Asymptotics  and the $S$-Matrix}

While much of the high energy physics community was gathered in Kiev, 
I was moving from Cambridge, where I had just completed my Ph.D., to Geneva, to begin a postdoctoral fellowship. I already knew CERN from the marvelous summer I had spent there three years earlier as a vacation student, assigned to the $g-2$ experiment, on the anomalous magnetic moment of the muon. It had been an extraordinary privilege and education to participate in this beautiful experiment as a very junior temporary colleague of Emilio Picasso, Leon Lederman and others. 

Returning to Cambridge  in October 1967 after my experimental vacation at CERN, I had started research in theoretical high energy physics under the supervision of John Polkinghorne. The prevailing intellectual ethos is well reflected in the book, {\it The Analytic $S$-Matrix} \ELOP, which Eden, Landshoff, Olive and Polkinghorne had produced the previous year. I, and nearly all my fellow research students, worked on strong interaction physics. (One of us was trying to work out the correct Feynman rules for gauge field theories, but this tended to be regarded as a rather recondite or eccentric enterprise.) 

The general strategy was to seek to determine and elucidate the analyticity properties of the $S$-matrix, {\it i.e.} the singularity structure implied and determined by unitarity, in the expectation or hope (depending on the strength of one's faith) that the particle spectrum and the scattering amplitudes might be determined, with some minimal number of further assumptions, by some sort of `bootstrap' process. It seemed that ideas related to analyticity would need to be supplemented by assumptions about high energy behavior,  described in terms of Regge theory. 

The guide for analyzing both analyticity structure and high energy behavior was perturbative quantum field theory, usually a simple $\phi^3$ scalar field theory, to avoid ``the inessential complications of spin''. 
It was not that anyone in Cambridge, in those days before asymptotic freedom,  thought that perturbative quantum field theory would be directly relevant to strong interaction physics (as far as I know), rather that it provided a {\it  formal} solution to the unitarity conditions and, in a certain sense, exhibited Regge high energy behavior.  Thus it provided a guide as to what to assume or what to try to prove in the more abstract context of $S$-matrix theory; it was a sort of theoretical laboratory. 

The analyticity structure of the $S$-matrix in the physical region was already well understood: the singularities occur on curves specified by the Landau equations with discontinuities specified by Cutkosky rules. My own first steps were in the direction of trying to elucidate aspects of the singularity structure outside the physical region. Feeling somewhat isolated, after about eighteen months, I moved to work on Regge theory, collaborating with my contemporary Alan White, following the group theoretical approach pioneered by Marco Toller, using harmonic analysis on noncompact groups. 

Toller was a staff member in the Theory Division at CERN at the time and, encouraged by David Olive, who was spending a couple of terms on leave there, he arranged a two-year postdoctoral position for me. David was then a Lecturer in DAMTP, Cambridge, and, when he returned from leave in May 1970, he gave a course of lectures on Dual Models, based in part on the work he had been doing with Alessandrini, Amati and Le Bellac at CERN.  Presented with his characteristic simplicity, elegance and depth,  the subject strongly interested me, because of the sense he conveyed of an emerging mathematical structure. 
Further, it offered a new theoretical laboratory, seemingly {\it somewhat} different from that of the Feynman diagrams of quantum field theory, with the Regge asymptotic behavior present from the tree diagrams onwards (rather than contained in certain selectively chosen infinite sums of diagrams as in quantum field theory), in which one could study the analyticity and high-energy properties of the $S$-matrix. It was this theoretical potential, rather than any phenomenological relevance, that engaged my interest. 

\newsec{Cuts and Poles}

Arriving at CERN in September 1970, I found an air of excitement surrounding these ideas, and number of physicists, mainly younger ones, working on them, inspired in large part by the leadership of CERN staff member Daniele Amati. Over the coming two years nearly everyone making fundamental contributions to dual models, or dual theory as it was sometimes more portentously called, passed through or came to stay. David Fairlie had come on leave from Durham, a little later Ian Drummond arrived from Cambridge and, in July 1971, David Olive returned to take a post as a staff member.

I began my own efforts by trying to understand the details of how the Veneziano model, together with the higher loop amplitudes that were being constructed, would provide a formal perturbative solution to the unitarity equations \FGW. There was, however, what looked like a fly in the ointment. In analyzing the divergences that occurred in one-loop dual amplitudes, which one needed to try to regularize by a renormalization process, a singularity in momentum had been found, coming from the same part of the integration region as the divergences.  This singularity was a cut, which it seemed would imply a violation of unitarity.

The one-loop amplitudes in the Veneziano models were associated with diagrams involving either an annulus (orientable case) or a M\"obius band (nonorientable case), to the edges of which were attached lines associated with the initial and final state particles and, so, incoming and outgoing momenta. In the string picture, which, as we have said, was used at the time as an analogue or figurative representation rather than a rigorous basis for calculation, the annulus or band corresponded to the world-sheet describing the history of the string or strings formed by the joining or splitting of the incoming string states. For the orientable case of the annulus, there were two sub-cases to consider: a planar case where all the momenta entered on the outside edge of the annulus; and a nonplanar case where the momenta entered on both edges. It was this last case that contained the unitarity-violating cut. 

The divergence in the integral corresponding to the annulus came as the radius of the circle constituting its inner edge tended to zero. In the nonplanar case, where momenta followed into both edges of the annulus, the integral corresponding to the loop amplitude could be made finite by taking the momenta entering on the inner edge to sufficiently unphysical (space-like) values and, from this region,  the amplitude could be defined more generally by analytic continuation. Further, it seemed that it should be possible to obtain a finite result for the planar case, from the nonplanar case, by analytically continuing all the momenta on the inner edge to zero, leaving only those on the outer edge.

Using this observation, I sought to develop this analytic approach to renormalization of dual loops in a paper written in February 1971 \Ga, but there was a problem: as the momenta on the inner edge were taken to zero, through real but space-like values, the onset of the divergence corresponded to a cut, the unitarity-violating cut to which we have referred. We could analytically continue past the cut to obtain a finite answer for zero momenta but, because of the cut, the value would inevitably be complex rather than real as desired, and the answer would depend on whether we continued above or below the singularity in the complex plane. 

On the other hand, if the theory could be modified so that the cut became a pole, unitarity would no longer necessarily be violated, provided that the theory contained new particles corresponding to this pole, and the regularization technique I was pursuing might well work. Moreover, as noted in the paper \Ga, Daniele Amati told me that Claud Lovelace had been considering such a possible modification of the theory. As we now know, in the `correct' version of the theory, the cut does indeed become a pole, corresponding to the propagation of a closed string, and the analytic renormalization procedure I was seeking to develop leads to an interpretation of the divergence of the planar loop in terms of the emission of closed string states into the vacuum, and its removal by a renormalization of the slope of the Regge trajectories \refs{\Shapa, \AAA}.

Lovelace's considerations were in fact probably the most prophetic -- perhaps most Delphic would be more appropriate -- in the development of string theory. He had recently left CERN, because his post as a staff member had not been made permanent, and he was now at Rutgers working on extracting a calculus of Reggeons and Pomerons from dual models to describe high energy scattering. The Reggeons corresponded to the external particles of the original Veneziano model and their excited states, whilst the Pomerons corresponded to the singularity (initially a unitarity-violating cut) that we have been discussing. Lovelace needed this singularity to be a pole for consistency with unitarity, of course, and, at the beginning of November 1970 \Lb, he started modifying the theory in an {\it ad hoc} fashion to achieve this.  He considered the possibilities that the dimension, $D$, of space-time might be different from $4$ and  that $E$ infinite sets of oscillators might be cancelled by `Virasoro-like' conditions, rather than just the one infinite set  that one might na\"\i vely expect. He found that he could remove the cut and obtain the desired pole by setting $D=26$ and $E=2$ \Lc.

\newsec{Searching for Ghosts}

At the same time as Lovelace, others were also considering what could be learnt from varying the space-time dimension, $D$. In the summer of 1970, Richard Brower and Charles Thorn had begun a study of the physical states of the theory. As we have recounted, Virasoro had found that, when $\alpha_0=1$, the physical states, $|\psi\rangle$,  satisfy an infinity of conditions,
\eqn\phys{
L_0|\psi\rangle=|\psi\rangle;\qquad L_n|\psi\rangle=0,\quad n>0,}
where $L_n$ is given by the normal ordered expression
\eqn\defL{
L_n={1\over 2}\sum_n:a_m\cdot a_{n-m}: .
}
[Note that $L_n^\dagger =L_{-n}$.] Fubini and Veneziano \FVb\ and Del Giudice and Di Vecchia \DD\ began the analysis of the Virasoro conditions \phys\ using the operator formalism \acomm, but Brower and Thorn \BT\  went further and set out a general algebraic approach to determining whether the conditions were sufficient to ensure the absence of the unwanted ghost states, characterizing the problem entirely in terms of the Virasoro algebra,
\eqn\Vir{
[L_m,L_n]=(m-n)L_{m+n}+{c\over 12}m(m^2-1)\delta_{m,-n},
}
where $c=4$ for the Veneziano model as originally formulated in four-dimensional space-time.

Virasoro \Virb\ had not actually written down the algebra that bears his name and the final $c$ term was initially omitted in discussions until Joe Weis (who tragically died in a mountaineering accident in 1978) pointed out that it should be there \refs{\FVb, \BT}. Brower and Thorn set about analyzing the norms of physical states by considering the signature of the orthogonal space, spanned by states of the form $L_{-n}|\phi\rangle$, called spurious states. Their profound paper \BT, now unfortunately not well  known, introduced {\it ad hoc} in their essence the concepts of a Verma module and the contravariant form (see {\it e.g.} \Wak).

They noted that the condition for the absence of ghosts  could be expressed in terms of the eigenvalues of the (contravariant form) matrices,  $\M_N(c,h)$, formed from the scalar products of the states
\eqn\verma{
L_{-1}^{n_1}L_{-2}^{n_2}\ldots L_{-m}^{n_m}|h\rangle, \quad\hbox{where } \quad
\sum_{j=1}^mn_j =N, }
the $n_j$ are non-negative integers and $|h\rangle$ is a normalized state satisfying 
$ L_0|h\rangle =|h\rangle$ and $L_n|h\rangle =0$ for $ n>0$. The condition was that, for $N>1$, $h+N\leq 0$, $h$ integral,  the signature of $\M_N(c,h)$ should be the same for as that for the matrices obtained by replacing $L_{-n}$ by $a_{-n}^0$ in the defining equations for $\M_N(c,h)$. 

In all this, $c$ equals the dimension of space-time, $D$, so that, naturally,  Brower and Thorn took $c=4$ in their analysis. They showed that their condition held up to the ninth level of states in the model. By about November \Thorna, they had started allowing $D$ to vary in their analysis, with the objective of establishing the result for $D=4$ by showing that no eigenvalues passed through zero as one continued from a region in which the result was true. Although this line of argument could not be completed at the time, it did lead to the discovery of a ghost amongst the physical states if the dimension of space-time, $D>26$,
as they noted in a footnote added to their paper \BT. (Some years later, Charles Thorn \Thornb\ gave a proof the absence of ghosts completing this original line of argument, using the famous determinant formula Kac had found for $\det\M_N(c,h)$ \refs{\Kaca, \FF}.)

Thus, early in 1971, the suggestion that 26-dimensional space-time had a particular significance in the context of dual models had come almost simultaneously from studies of two apparently unconnected aspects of the theory. This seemed deeply mysterious but hardly likely to be a complete coincidence. One thing was clear: the dimension of space-time had to play a role in any proof of the absence of ghosts.

I learnt about the results on ghosts directly from Rich Brower, who had arrived at CERN as a postdoctoral fellow about the same time as I had. We had began to collaborate, initially looking for new dual models. Soon after Veneziano had produced the first dual model, Virasoro had produced a model with somewhat similar properties \Virb. In fact, it turned out to be closely related to the original model. Shapiro \Shapb\ showed that the $n$-point Virasoro model could be obtained by a similar analogue approach but this time integrating the locations of the external momentum lines (Koba-Nielsen variables) over the whole Riemann sphere rather than just the circular boundary of a disk. Further, the model also had an infinite number of physical state conditions, which might well eliminate all the ghost states in suitable circumstances. 

Rich and I sought to generalize the Virasoro model by replacing the two-dimensional integral by an $n$-dimensional integral, with the group responsible for duality generalizing from the SO(2,1) that Gliozzi had found for the Veneziano model ($n=1$) and SO(3,1), which fulfilled the role for the Virasoro model ($n=2$), to SO($n+1$,1). But the problem was that, whereas the algebras of these groups have suitable infinite-dimensional extensions for $n=1, 2$ (being isomorphic to SL(2,$\Rop$) and SL(2,$\Cop$), respectively), this is not the case for general $n$. So our new models \BGa\  lacked the infinite-dimensional algebra (\Vir\ for $n=1$ and two copies of it for $n=2$) needed to have a chance of eliminating ghosts. 

Still we thought it worth doing some simple ghost searches, in part motivated by the scary rumor that a student had found a ghost coupling in the four-point Veneziano model at about the 13th level. We spent a weekend doing a computer-aided search  and found nothing to worry about down to the 30th level (for $1\leq n\leq 45$), which scotched the rumor and restored our confidence about the conjectured absence of ghosts in the original Veneziano model.

Meanwhile, very much more interesting models were emerging elsewhere. The model of Veneziano was proposed as a model for  mesons and had only bosonic states. In January, Pierre Ramond \Ram\ produced an equation describing a free dual fermion, generalizing in a sense the free Dirac equation. Soon after, Andr\'e Neveu and John Schwarz \NSa\ used fermion oscillators to construct a new dual model for mesons, which could interact with Ramond's fermions, and which possessed an infinite-dimensional superalgebra of physical state conditions \NST\  that offered the prospect of ghost elimination. 

It was an interesting, if quick and straightforward, task to calculate one-loop amplitudes for the Neveu-Schwarz model, so Ron Waltz and I did this exercise in June 1971, and it was indeed fun to see the predictable elliptic functions emerging for the first time \GW.  
For one thing, it enabled one to perform immediately the analysis that Lovelace had performed for the Veneziano model and determine the value of the space-time dimension, $D$, for which the unitarity-violating cut in the nonplanar loop became an acceptable pole, the answer being $D=10$ instead of $D=26$. I do not remember at what point someone did a calculation, analogous to that of Brower and Thorn, to show that the model possesses ghosts for $D>10$, so that it became clear that $D=10$ plays exactly the role for the Neveu-Schwarz-Ramond (NSR) model  that $D=26$ does for the Veneziano model.

\newsec{Spectrum Generating Algebra}

Thus the challenge was to prove the absence of ghosts in the Veneziano model for $D\leq 26$ and in the NSR model for $D\leq 10$, with the limiting or {\it critical} values of the space-time dimension being required for consistency with unitarity at the one-loop level. In June 1971, Del Giudice, Di Vecchia and Fubini (DDF) \DDF\ made a major advance in understanding the physical states by seeking to construct a basis of physical states, and hoping to prove their positivity, rather than just attempting to prove the absence of ghosts. Their approach was to consider the operator describing the coupling of the massless spin one particle, which exists in the theory under the assumption that $\alpha_0=1$, and might more naturally be identified, at least figuratively, with the photon than a vector meson (belying the interpretation of the dual model as a theory of strong interactions). 

To describe the DDF operators, we need to select a light-like vector, $k$.
DDF worked in four-dimensional space-time but, with a view to subsequent developments, we consider general $D$-dimensional space-time. The DDF operators are 
\eqn\ddfo{
A^i_n=\epsilon_\mu^iA^\mu_n, \qquad A^\mu_n={1\over 2\pi i}\oint P^\mu(z)V(nk,z)dz
}
where the $\epsilon^i$, $1\leq i\leq D-2$, are orthonormal polarization vectors for a photon of momentum $k$, {\it i.e.} $\epsilon^i\cdot\epsilon^j=\delta^{ij}$, $k\cdot\epsilon^i=0$,  and $n$ is an integer. In \ddfo, $V$ is the basic vertex operator,
\eqn\Vz{
V(\alpha,z)= \ :\exp\left\{i\alpha\cdot Q(z)\right\}:\ 
}
where
\eqn\Qexp{
Q^\mu(z)=q^\mu-ip^\mu\log z +i\sum_{n\ne 0} {1\over n}a^\mu_nz^{-n}, \qquad
P^\mu(z)=i{dQ\over dz}^\mu,
}
with $a_0^\mu=p^\mu$, the momentum operator, and $[q^\mu,p^\nu]=ig^{\mu\nu}$.
The DDF operators, $A^i_n$, are well-defined on states of momentum $p$, such that $k\cdot p=1$
and they can be used to generate physical states because they commute with the Virasoro algebra,
\eqn\VLc{
[L_m,A^i_n]=0.
}

DDF showed that the $A^i_n$ create an orthonormal set of positive norm physical states corresponding to $D-2$ dimensions of oscillators.  But, to get all the physical states \phys, we need $D-1$ dimensions of oscillators. 

I started studying with Rich Brower the progress DDF had made. We focused more directly on the algebraic properties of the DDF operators, noting that  they really did behave like annihilation and creation operators, satisfying the appropriate commutation relations \refs{\BGb,\BGc},
\eqn\ddfcom{
[A^i_m,A^j_n]=m\delta^{ij}\delta_{m,-n},\qquad 1\leq i,j\leq D-2.
}
We set out to find the operators that would create the remaining physical states, with the hope that we could then prove the absence of ghosts using the algebraic properties of these additional operators. 

Because the operators $A^\mu_n$ are associated with the coupling of a ``photon'', one would not expect the components other than those in the directions of the polarization vectors $\epsilon^i$ to be physical. There are two other independent components, which we can take to be in the direction $k$ of the ``photon'' momentum and in a longitudinal direction defined by a light-like vector $\tilde k$ satisfying $k\cdot \tilde k=-1$. The $k$ components essentially vanish: $k\cdot A_n=k\cdot p\delta_{n,0}$ and the longitudinal components, $\tilde  A_n=\tilde  k_\mu A^\mu_n$, fail to commute with the Virasoro algebra,
\eqn\LAc{
[L_m,\tilde  A_n]=-\half m(m+1)n\Phi^n_m, \qquad
\Phi^n_m={1\over 2\pi }\oint e^{ink\cdot Q(z)}z^m{dz\over z},
}
though they satisfy a neat Virasoro algebra themselves \refs{\BGb,\BGc},
\eqn\ddfgb{
[\tilde A_m,\tilde  A_n]=(m-n)\tilde  A_{m+n}+m^3\delta_{m,-n}.}
This suggested that we should seek a modification of the $\tilde A_n$ by some function $F_n$ of the 
$K_m=k\cdot a_m$, 
\eqn\tAFL{
A^L_n=\tilde A_n +F_n,}
which would adjust $\tilde A_n$ so that the resulting longitudinal  operator $A^L_n$ commutes with $L_m$. Rich and I began to look for $F_n$.

Daniele Amati and Sergio Fubini arranged for me to talk about this work in progress at a summer school I was going to into Varenna in August  \BGb. An impressive list of lecturers gathered in that beautiful setting on Lake Como, including Callan, Dashen, Coleman, Goldberger (who was to be my predecessor but one as Director of the Institute for Advanced Study), Maiani, Regge and Salam, as well as Amati and Fubini.  

For me, it was a memorable meeting and one particular vignette has stuck in my mind as an illustration of the prevailing attitude towards the use of modern mathematics in theoretical high energy physics. A senior and warmly admired physicist gave some lectures on the Regge theory of high energy processes. With great technical mastery, he was covering the board with special functions, doing manipulations that I knew from my studies with Alan White (who was also at the School) could be handled efficiently and elegantly using harmonic analysis on noncompact groups. Just as I was wondering whether it might be too impertinent to make a remark to this effect, the lecturer turned to the audience and said, ``They tell me that you can do this all more easily if you use group theory, but I tell you that, if you are strong, you do not need group theory.''

Rich and I continued our search for the modification $F_n$ together in Varenna and on a subsequent visit to Cambridge, and at a distance after Rich returned to the US to resume a postdoctoral fellowship at MIT, and I returned to CERN for my second year. But before I went back to Geneva, I went to Durham to be interviewed for a faculty position in the Department of Mathematics. I had enjoyed discussions with David Fairlie, who had been on leave at CERN from Durham, however I did not really think I stood much chance of getting the job, because I was only one year postdoctoral. 

Just a couple of years earlier this would not have been seemed an obstacle. Indeed, I remember that when I started research, it was explained to me that a recent doctoral graduate had gone straight to an essentially tenured post at a good UK university because, since he was good at teaching but not research, there was no point his spending time in a postdoctoral research post.  But now things were getting very much more difficult; John Ellis estimated a lower bound of 50 for the number of UK theoretical physicists looking for university posts.  When I was called for interview, I was convinced that they would appoint the other, more senior, candidate. Perhaps because of this I was not nervous, and perhaps consequently got the job. Fortunately, Durham agreed that I could go back to CERN to complete the second of my two years there. Having a reasonably secure future, I could concentrate on my research. 

By September 1971, Rich was back in  the United States and we continued our collaboration by post. I worked on some other ideas as well but my mind kept grappling with trying to understand the physical states and prove the absence of ghosts. My wife, Helen, and I had {\it abonnements} at the Grand Th\'e\^atre in Geneva, which ensured we saw ten operas and ballets  a year, but I fear it was often the structure of physical states that I was more conscious of than the pattern of dancers on the stage. At the beginning of November, Rich and I submitted our paper \BGc\ on the spectrum generating algebra for the dual model to Nuclear Physics, still lacking a form for $F_n$ to make the correct physical states creating longitudinal operators, $A^L_n$.
 
 \newsec{No-Ghost Theorem}
 
In January 1972,  Charles Thorn arrived at CERN from Berkeley to spend a year and, after a while, we began to discuss our individual efforts to prove the absence of ghosts. Charles had been studying further the physical states as a function of the space-time dimension, $D$. Brower and Thorn \BT\  had noted that a ghost state appears amongst the physical states when $D>26$. The critical dimension, $D=26$, marks a transition point at which the norm of a physical state moves from positive values through zero to negative values.

In each dimension $D$, states of the form $L_{-1}|\phi\rangle$ are null physical states, {\it i.e.} physical states orthogonal to all other physical states, provided that $|\phi\rangle$ satisfies the physical state conditions $L_n|\phi\rangle =0, n>0$, with the adjusted mass condition $L_0|\phi\rangle=0$. Exactly for $D=26$, we have more  states like this:
\eqn\nullst{
\left( L_{-2} + {\scriptstyle {3\over 2}} L^2_{-1}\right)|\phi\rangle}
are null physical states for any $|\phi\rangle$ satisfying $  L_n|\phi\rangle=0, n>0,$ and
$ L_0|\phi\rangle=-|\phi\rangle$. They correspond to a zero eigenvalue of $\M_2(D,-1)$ at $D=26$, and this makes a transition to a negative eigenvalue, corresponding to a ghost, for $D>26$.  Simple counting shows that the null physical states (6.1) cannot all be independent of those
of the form $L_{-1}|\phi\rangle$, because, if they were, the null physical states taken together with the DDF states would 
exceed the known number of physical states. So there must be relations between them, corresponding
to further zero eigenvalues of the matrices $\M_N(D,h)$ for $D=26,N>2$ and  $h\leq 1-N$. From his detailed calculations, Charles began to uncover  an intricate structure of such relations 
at $D=26$.

Suddenly  the realization came that perhaps all the physical states, other than the DDF states, which we knew to be positive definite,  became null at $D=26$. As I remember, Charles and I were walking into the CERN cafeteria to have lunch when  lightning struck. This would mean that, at $D=26$, only the transverse DDF modes would be needed to construct the physical states; the additional transverse modes $A^L_n$ that Brower and I had been seeking would not correspond to physically distinct modes of motion in the critical dimension. 

Moreover, this would make the prophecy of Lovelace come to pass, for he had written \Lc\ that not only should we have $D=26$ but also $E=2$, that is that two dimensions  of oscillators should be effectively cancelled by the Virasoro conditions \phys, because the $D-2=24$ dimensions of oscillators described by the DDF operators create all the physical states apart from the null states that we would expect to drop out of the theory. If they do indeed decouple from loops, the unwanted cuts should become poles consistent with unitarity.

Charles set about trying to find recursively all the null states necessary to establish the accuracy of this picture and so prove the absence of ghosts. Then we thought it might be easier to seek to characterize the DDF states algebraically and show that any physical state could be written as the sum of DDF state and a null physical state. The space generated by the DDF states can be specified by imposing the conditions
\eqn\ddfk{K_n|\psi\rangle=0,\qquad n>0,}
in addition to the physical state conditions \phys. The $K_n$ together with the $L_n$ form an extension of the Virasoro algebra, 
 \eqnn\LKalg
 $$\eqalignno{
 [L_m,L_n]&=(m-n)L_{m+n}+{D\over 12}m(m^2-1)\delta_{m,-n}\cr
 [L_m,K_n]&=-nK_{m+n}&\LKalg\cr
 [K_m,K_n]&=0.
 }$$
 We also define the space of {\it null physical states} to be the states satisfying \phys\ that are linear combinations of states of the form $L_{-n}|\phi\rangle$, $n>0$. Such states are obviously orthogonal to all physical states. 
Having formulated the problem like this, we were able to prove the No-Ghost Theorem \GT:  For $D=26$, if $|\psi\rangle$ is a physical state, {\it i.e.} a solution of \phys, it can be written in the form
\eqn\ngthm{
|\psi\rangle = |f\rangle + |n\rangle,}
where $|f\rangle$ is a DDF state and $|n\rangle$ is a null physical state. The No-Ghost Theorem follows for $D<26$ by embedding in the $D=26$ case. We were also able to prove the No-Ghost Theorem for the Neveu-Schwarz dual model and the Veneziano model at the same time by essentially the same argument, confirming the critical dimension as $D=10$ for the NSR model.

As we were working towards our proof, we heard indirectly that Rich Brower had completed a proof of the absence of ghosts. Earlier, in January, soon after Charles had arrived in CERN, I had heard from Rich that he was planning to write a paper about  ideas he was working on for proving the No-Ghost Theorem, progressing  further along the route we had been following, but that he then still lacked a construction for $F_n$. Charles and I thought we should publish our proof, because it was conceptually different and because it had been arrived at independently, but we did not want to do anything which might have the effect of scooping Rich, so we delayed circulating our paper for three weeks until the beginning of May and we adopted a low key, perhaps slightly obscure, title, {\it The Compatibility of the Dual Pomeron with Unitarity and the Absence of Ghosts in the Dual Resonance Model}.

To frame his proof, Rich had realized that if the addition of $F_n$ to $\tilde A_n$ to form $A^L_n$ were to double the $c$ number in   \ddfgb, so that
\eqn\ddfL{
[A_m^L, A_n^L]=(m-n) A^L_{m+n}+2m^3\delta_{m,-n},}
then, for $D=26$, there would be in essence an isomorphism between, on the one hand, the algebra of the $A^i_n$ and $A^L_n$ and, on the other, the algebra of Euclidean oscillators, $a^i_n, 1\leq i\leq 24$ and the Virasoro generators formed from them, $L_n^a$. Assuming that the $A^i_n$ and $A^L_n$ did create all the physical states,  the isomorphism of the algebras could be used to calculate their norms using the operators $a^i_n$ and $L^a_n$ instead,  and, since these act within a space with no negative norm states, there could be no ghosts.

Rich found the $F_n$ we had been seeking \Bro,
\eqn\defF{
F_n={n\over 4\pi i}\oint {k\cdot P'(z)\over k\cdot P(z)}V(nk,z)dz,
}
for which it is indeed the case that $A^L_n=\tilde A_n+ F_n$ satisfies \ddfL\ and 
\eqn\LAL{
[L_m,A^L_n]=0, \qquad [A^L_m, A^i_n]=-nA^i_{m+n}.}
Then, for $D=26$, the algebra of the $A^i_n, A^L_n$ is essentially isomorphic to that of the $A^i_n, L_n^A$, where the $L^A_n$ are Virasoro algebra formed from the $A^i_n$.  It follows that the operators $A^L_n-L^A_n$ 
commute with the $A^i_n$ and it is easy to show that they create null states, leading to the same conclusion \Bro\ about the structure of the physical states as in \ngthm. 

This approach to proving the No-Ghost Theorem was extended by John Schwarz \Sch\ to the Neveu-Schwarz model and this was also done by Rich and Kenneth Friedman \BF.

The validity of the No-Ghost Theorem had a profound effect on me. It seemed clear that this result was quite a deep mathematical statement, and that the proofs involved elegant argument and revealed a beautiful structure, but also that no pure mathematician would have written it down. It had been conjectured by theoretical physicists because it was a necessary condition for a mathematical model of particle physics not  to be inconsistent with physical principles. Of course, it is reverse logic and perhaps dubious philosophy, but I could not help thinking that, in some sense, there would be no reason for this striking result to exist unless the dual model had something to do with physics, though not necessarily in the physical context in which it had been born.

The deep mathematical significance of the No-Ghost Theorem was demonstrated two decades later when it played an important role in Borcherds' proof \Bor\ of Conway and Norton's  ``moonshine conjectures'' about the Monster Group. I felt enormously flattered, but faced the most formidable challenge,  when I was invited to deliver the laudation \Gicm\ for Borcherds at the International Congress of Mathematicians in Berlin in 1998, when he was awarded a Fields Medal.

\newsec{Quantizing the Relativistic String}
  
Claudio Rebbi, also at CERN on a postdoctoral fellowship, had been studying the physical states of the dual model by considering the limit in which the momentum of the states became infinite in a particular Lorentz frame. Charles and I began investigating with Claudio what happened to the DDF states when a rotation is made about the momentum of the state. If $E_{Li}$ is the generator of the little group of the momentum that generates a rotation in the $i$-th longitudinal plane, we found
\eqn\Lirot{
[E_{Li}, A^j_m]=\delta^{ij}A^L_m+\sum_{n=1}^\infty {1\over n}\left(A^i_{-n}A^j_{m+n}-A^j_{m-n}A^i_n\right)
}
with $A^L_n$ as defined by Brower \Bro. Actually, \Lirot\ can be taken as a definition of $A^L_n$ and it can be shown \GRT\ then that the properties of $A^L_n$, namely \ddfL\ and \LAL, follow from this definition, avoiding the need to construct $F_n$,  so that this provides a construction of all the physical states and a basis for Brower's proof of the absence of ghosts. A similar argument can be used in the Neveu-Schwarz model. 

Now, because of the No-Ghost Theorem, we know that, in the critical dimension $D=26$, the DDF operators generate all the physical states, provided that we disregard the null states, which do not contribute to tree amplitudes. Thus we must be able to define the action of all the generators of the little group of the momentum, including $E_{Li}$, on the transverse states generated by the DDF operators.
We found the corresponding expression for $E_{Li}$,
\eqn\LirotLA{
E_{Li}=\sum_{n=1}^\infty {1\over n}\left(L^A_{-n}A^i_{n}-A^i_{-n}L_n^A\right)
}
and noted that so defined $E_{Li}$ closes with $E_{ij}$, the generators of rotations in the $(i,j)$-plane, to form the SO(D-1) algebra of the little group, if and only if $D=26$.

In the limit $k\rightarrow 0$ (and $\tilde k$ becomes infinite), the DDF operators $A^i_n$ become the basic oscillators  $a^i_n$, the basic oscillators of the theory (suitably identifying axes), and the $L^A_n$ become $L^a_n$, the Virasoro generators made from $a^i_n$.  This suggests that, in a suitable limit, the description of the dynamics of the system might simplify. This took us back to reconsidering the attempts to understand the dual  model in terms of strings.

As outlined in Section 1, very soon after the oscillator description of the dual model had been obtained \refs{\FGV, \Nama}, it was realized that this implied that the model could be regarded as describing some sort of physical system specified in terms of a one-dimensional object or string \refs{\Nama, \HBN, \Sc}. Initially, the main impact of this picture was through the analogue approach introduced by Nielsen \HBN, because it suggested the appropriate mathematical techniques for loop calculations \refs{\Loa, \Alessa}. It was unclear whether this should be regarded merely as an analogy and calculational guide or as a deeper physical description. 

In 1970, Nambu \Namb, and then Goto \Goto, had written down an action \LNG\ that would prove suitable for describing the dynamics of the string, but they had not then developed the analysis of this action to derive its properties. Nambu's paper was not widely circulated and the Nambu-Goto action, as it later came to be called, was little discussed. I remember Jeffrey Goldstone, who had come on leave from Cambridge to CERN for the first part of 1971, writing down this action and asking me whether it was what people were talking about when the string picture of the dual  model was referred to. I replied that I wasn't sure.

I had known Jeffrey since I had been a freshman at Trinity College, Cambridge, where he was then a Fellow. As an undergraduate, I had attended courses of lectures by him on theoretical physics. These did not always draw large audiences but I found them deeper than some other offerings. Certainly, there was the sense of the subject being developed in real time rather than something prepared earlier being warmed up. (Jeffrey claimed to me years later that the folded notes, produced from his pocket at the beginning of each lecture, on which it was apparently based, were in fact a stage prop.) 

When I was a graduate student in DAMTP, Jeffrey was not much in evidence in the department. A few times I went to seek him out for advice in his rooms in Trinity College, where I noticed a rather attractive representation of a hippopotamus from the Ny Carlsberg Glyptotek in Copenhagen. Jeffrey combined a lugubrious demeanor with a penetrating wit. I remember him counseling that the nature of research was not so much the answering of predetermined questions as ``What is the precise question to which this is the vague answer?" At CERN, there was a visitor, presumably unofficial, dressed in white, wearing gold trainers, who came to seminars. It seemed he originated from the World Questions Institute, apparently a private house boat on the Hudson River.  He was looking for questions rather than answers. He sat at the back of seminars, feeling the vibrations, which  were most stimulating during talks on the ghost problem in dual models it seemed. 

Early in 1972, building on earlier work of Chang and Mansouri \CM, Mansouri and Nambu \MN\ noted that the geometric character of the action \LNG\  implies that it is independent of the choice of coordinates 
$(\sigma, \tau)$, and then
showed that the Virasoro conditions \phys\ were related to  a choice of coordinates that ensures the equality of the Lagrangians $\L_O$ of \quadratic\ and  $\L_{NG}$ of \LNG.

As Charles, Claudio and I sought to understand how the covariance properties of the states of the dual  model, which we had obtained \GRT,  related to the quantization of a string described by the Nambu-Goto action, we heard through David Fairlie of work that Jeffrey Goldstone was doing back in Cambridge on the same topic. When we wrote to him at the end of July to  exchange information about our work, his reply began characteristically: ``My questions are the same as your questions and I have no answers'', before giving a succinct summary of the progress he had made, that overlapped considerably with what we had been doing.  He explained that he intended to write up his results but he had ``been chasing a different hare, surfaces instead of strings, these past two weeks''. 

Following Chang, Mansouri and Nambu \refs{\CM, \MN}, we made a choice of coordinates $(\sigma,\tau)$ on the world sheet in which
\eqn\ortho{
\left({\partial x\over\partial \sigma}\right)^2+
\left({\partial x\over\partial \tau}\right)^2=0,\qquad 
{\partial x\over\partial \sigma}\cdot{\partial x\over\partial \tau}=0.}
Such an {\it orthonormal } choice of coordinates not only ensures that $\L_{NG}=\L_O$ (for the particular solution) but it also linearizes  the equations of motion that follow from variation of $\A_{NG}$, 
removing the nonlinearity that had initially deterred Nambu \Namb. In an orthonormal coordinate system these nonlinear equations are equivalent to the wave equation for the world sheet variable $x^\mu(\sigma,\tau)$. 

This enables the introduction (at the classical level) of the harmonic oscillator variables $a^\mu_n$ to describe the normal modes of the solutions of the wave equation,
\eqn\xexp{
{1\over \ell}x^\mu(\sigma,\tau)=q^\mu+p^\mu\tau+i\sum_{n\ne 0}\,
{a^\mu_n\over n}e^{-in\tau}\cos n\sigma.
}
[Here the dimensional parameter $\ell$ is  the characteristic string length, which will be specified in (7.5).] In terms of these variables, the conditions \ortho\ are equivalent classically to the vanishing of the Virasoro  generators, $L_n$ \refs{\CM, \MN}.

It is possible to specialize further our choice of orthonormal world-sheet coordinate system by choosing $\tau$ to be the time coordinate in some Lorentz frame, $\tau=\n\cdot x$, where $\n$ is a time-like vector. Then $\n\cdot a_n=0, n\ne 0$, so that the time coordinates $x^0(\sigma,\tau)$ do not describe independent oscillations. This removed an obscurity in the original space-time description of the string, about what the apparent oscillations of the string in the time direction meant. 

Further, the classical equations $L_n=0$, corresponding to \ortho, allow a further component of $a^\mu_n$ to be eliminated in principle, leaving a set of independent degrees of freedom corresponding to $D-2$ dimensions of oscillators. Of course, this is what one should expect from the geometric properties of the Nambu-Goto action, $\A_{NG}$. It only depends on the area of the world sheet so that changes of $x^\mu(\sigma,\tau)$ tangential to the world sheet will not change the action and are not dynamically significant; only the $D-2$ transverse dimensions of oscillations are significant. 

If the string picture of dual models, with dynamics specified by the action, $\L_{NG}$, suggested by Nambu \Namb, had been understood earlier, the path to understanding the structure of the physical states and the proof of the No-Ghost Theorem would have been clearer. For the degrees of freedom in the quantum theory to correspond to those in the classical theory ({\it i.e.} only transverse oscillations are physical),  the Virasoro conditions must effectively eliminate two dimensions of oscillators. If this basic aspect of string dynamics had been understood, Lovelace's prophecy that $E$, the number of dimensions of oscillators removed, should be 2, as well as $D=26$, would have seemed somewhat less obscure; and it would have been evident that the DDF construction  \DDF\  should provide all the physical states, as it does in the critical dimension $D=26$, pointing the way to the proof of the No-Ghost Theorem. In fact, this was all only evident with hindsight. 

A difficulty with solving the classical constraint equations $L_n=0$, after substituting $\n\cdot a_n=0, n\ne 0$, is that the equations are quadratic. This difficulty is removed if $\n$ is taken to be light-like rather than time-like, $\n=k$, because then $k\cdot a_n=0, n\ne 0$ and the constraint equations $L_n=0$ express $\bar k\cdot a_n$ in terms of the transverse components $a^i_n$ of the oscillators: $\tilde k\cdot a_n = L^a_n, n\ne 0$, the Virasoro generators made from these transverse components. [Here $\tilde k^2=0, \tilde k\cdot k=-1, k\cdot p=1$, as in Section 5.]

Using this light-cone approach, we canonically quantized the string, obtaining a spectrum built from only  $D-2$ dimensions of transverse oscillators \GGRT. The disadvantage with this approach is that the solution of the constraint conditions \ortho\ has been achieved at the cost of the choice of $k$, which breaks manifest Lorentz invariance. To restore Lorentz invariance to the resulting theory, one must specify how to perform Lorentz transformations, particularly those generated by the  $E_{Li}$. The straightforward expression for these is given by \LirotLA\ with $A^L_n$ replaced by $\tilde k\cdot a_n=L^a_n$ and $A^i_n$ replaced by $a^i_n$. But, we knew from our previous work \GRT\ that these expressions only close to form the appropriate Lorentz algebra if $D=26$; otherwise there is an anomaly term. It is only for $D=26$ that the canonical quantization procedure is Lorentz invariant. Thus, the critical dimension condition was obtained from the process of canonically quantizing the relativistic string using light-cone coordinates.

[One could seek to modify the generators \LirotLA\  iteratively by terms that vanish in the classical limit (and thus do not violate the correspondence principle) so as to cancel out the anomaly terms and restore covariance for $D\ne 26$. The problem  is that, whilst this may be possible for the free string, it is not clear that it would be possible to introduce interactions satisfactorily, unless the procedure had a suitable geometrical basis. The results from dual models tend to suggest that a satisfactory theory could not be obtained in this way.]

An alternative approach to quantization, which preserves covariance at the sacrifice of being canonical, is to relax the constraints and treat all the oscillators $a^\mu_n$ as independent, imposing the constraints as weak conditions on the physical states in the quantum theory, {\it i.e.} the expectation values of \ortho\ in physical states should vanish as $\hbar\rightarrow 0$. This approach is analogous to the Gupta-Bleuler method of quantizing electrodynamics, but electrodynamics is relatively straightforward because the Lorentz gauge constraint is linear whereas \ortho\ is quadratic. The theory of quantizing constrained systems to handle this was developed in the 1950s by Paul Dirac \Dirac.

In this covariant approach, the conditions on the physical states take the form \phys\ and we knew from the No-Ghost Theorem \refs{\Bro, \GT} that these only define physical states free from ghosts if $D\leq 26$. Further, unless $D=26$, the states correspond broadly speaking to $D-1$ dimensions of oscillators rather just the $D-2$ transverse dimensions that would correspond directly to the classical theory.  
For $D<26$, although there are no negative norm states, there are `additional' states containing longitudinal oscillations of the string present in the quantum theory that do not have classical analogues.  

So, unless $D=26$, in the canonical approach there is a breakdown of Lorentz invariance and, in the covariant approach, anomalous longitudinal modes of oscillation are present in the quantum theory. Lovelace's calculation \Lc\ showed that these extra modes would result in unitarity-violating cuts when interactions were introduced and loop contributions calculated (unless some way of modifying the theory  could be found). For $D=26$, the noncovariant canonical approach and the noncanonical covariant approach to quantizing agree, and give a consistent way of quantizing the relativistic string, yielding precisely the elegant spectrum found in dual models. Subsequently, Stanley Mandelstam \Ma\ showed that interactions could be introduced in a simple way, by just allowing strings to split and to join at their ends. In this way, the original amplitude of Veneziano \Vena\ was obtained rigorously from a simple string picture. 

The cost of all this was the seemingly unphysical requirement that the space-time dimension $D=26$
(and the tachyon that results from $\alpha_0=1$).  But the critical dimension in the NSR model was $D=10$ and the general view at the time was that one should search for a `correct' model, with the sort of physical state structure that had been elucidated but with $D=4$ and no tachyons.

In order for the quantum commutators of the oscillators $a^\mu_n$ to be normalized as in \acomm, the characteristic length $\ell$ must be given by
\eqn\defell{
\ell=\sqrt{\hbar c\over T_0\pi}.}
With some ironic amusement, Jeffrey pointed out that, for the quantum states of the string to have masses comparable with those of hadrons, the tension of the string, $T_0$,  should have the somewhat macroscopic value of about 13 tons. The states on the leading Regge trajectory correspond to motions in which the string rotates like a rigid rod with the ends moving at the speed of light. Not everyone found our description of even the classical motion of the relativistic string plausible. One author characterized it as ``{\it r\'esultat \'evidemment faux}''  \JMS. 
Jeffrey's response was that it might conceivably be {\it faux} but it was surely not {\it \'evidemment faux}.

At the beginning of September 1972, I left CERN to take up my lectureship in Durham. On 28 August, just before I left, I gave a seminar on our work on the quantum mechanics of a relativistic string. The theorists working on dual models had been meeting at 2 pm most Thursdays throughout the academic year in a small discussion room, sitting on chairs round the periphery. The seminars were informal and there was a very free exchange of ideas. Amongst the regular participants at various times were Victor Alessandrini, Daniele Amati, Lars Brink, Edward Corrigan, Paolo Di Vecchia, Paul Frampton, David Olive, Claudio Rebbi, Joel Scherk and Charles Thorn.  Evidently news of our work in progress must have spread -- or perhaps it was  just because Murray Gell-Mann, who was visiting CERN that year, had expressed interest in it -- because a much larger audience than usual turned up and we had to relocate to the main theory conference room.

As we drafted our paper on the {\it Quantum Dynamics of a Massless Relativistic String} \GGRT, with me in Durham and  Charles and Claudio still in CERN,  we sought to persuade Jeffrey to join us in writing it, and eventually, shortly before it was produced at the end of October, he agreed. By then I was immersed in teaching my first lecture course, first-year applied mathematics.  The Head of Department, Euan Squires, told me that I should spend my initial lecture teaching them the Greek alphabet but, in spite of Res Jost's comment that, in certain periods, the only mathematics required of a theoretical physicist had been a rudimentary knowledge of the Latin and Greek alphabets \SW, I couldn't bring myself to do this and told them where to look it up. 

The two years I had spent in CERN had built up to an crescendo of intellectual excitement and, though I have found much of my subsequent research work gripping and often extremely satisfying (when teaching duties and the largely self-inflicted wounds of administration have permitted), nothing has quite matched this period. In particular, I had the privilege of working closely for seven or eight months with Charles Thorn, whose combination of deep perception and formidable calculational power had provided the basis of what we managed to do.
And, the exhilarating combination of the open and cooperative atmosphere that prevailed amongst (almost all) those working on dual models in CERN, the relative youth of most of those involved, the sense of elucidating a theory that was radically different, even the frisson of excitement that came from doing something that was regarded by some of those in power as wicked, because it might have nothing directly  to do with the real world -- this cocktail would never be offered to me again. 

\newsec{Fermions and Fast Forward}

I arrived at Durham at the same time as Ed Corrigan, who had just completed his doctorate in Cambridge  under the supervision of David Olive. He had been working with David on developing the understanding of amplitudes in the NSR model involving fermions. The model was then not yet complete because only tree amplitudes involving no more than two fermions (one fermion line) were known. 

As a step towards constructing amplitudes for four fermions ({\it i.e.} fermion-fermion scattering) and more, the vertex describing the emission of a fermion, so converting a boson line into a fermion line, had been constructed \refs{\Thornc-\CO}. We worked on understanding the gauge properties of this vertex \CGa\ with a view to ensuring that the fermion-fermion scattering amplitude would only involve the exchange of physical boson states -- otherwise the fermion-anti-fermion scattering might produce ghost boson states. 

The calculation of the tree amplitude for fermion-fermion scattering in the operator formalism was comparable in complication with the calculation of loop amplitudes in the bosonic theory of Veneziano. In loop amplitudes, care had to be taken in the operator formalism that had been developed to ensure that ghost states do not circulate round the loop. As I mentioned in Section 6, the No-Ghost Theorem shows that the spectrum of physical states in the critical dimension agrees with the counting, corresponding to $D-2$ dimensions of oscillators,  that Lovelace had deduced was necessary for loop amplitudes to be consistent with unitarity. 

But, that did not in itself constitute a construction of the bosonic loop amplitude with only physical states circulating and agreeing with Lovelace's conjectured formula \Lc;
this was achieved by Lars Brink and David Olive who constructed a projection operator onto the space of physical states \BO, providing incidentally another proof of the No-Ghost theorem. Using the projection operator of Brink and Olive, one could also ensure that only physical bosonic states were exchanged in fermion-fermion scattering. The net effect of doing this was to include a factor given by the determinant \OS\
\eqn\defDelta{
\Delta(x)=\det\left(1-M(x)^2\right),}
where the matrix
\eqn\Mmatrix{
M_{mn}(x)=(-x)^{\half(m+n+1)}{n+\half\over m+n+1}\left({-\half\atop m}\right)\left({-\half\atop n}\right).}
John Schwarz was particularly interested in the behavior of $\Delta(x)$ near $x=1$, because this governed the spectrum in the dual channel and  he had suggested \Scha\ that the Neveu-Schwarz model should describe gluon states and the Ramond sector quark states; then $q \bar q$ states dual to the gluon states should describe the meson spectrum, which might be interestingly different from the original Neveu-Schwarz gluon spectrum. By numerical computer calculation, Schwarz and Wu \SWu\ found that the leading behavior of $\Delta(x)$, which determined the lowest mass in the dual channel, was $(1-x)^{1\over 4}$. Indeed, they found  that, within the limits of their numerical calculation, 
\eqn\Deltaexp{
\Delta(x)=(1-x)^{1\over 4},}
exactly. 

This is how things stood when I reviewed dual models \Grev\ at the Aix conference in September 1973. Just  after the meeting,  Ed Corrigan and I produced a proof \CGc\ of the absence of ghost states coupling in the fermion (Ramond) sector of the RNS model, provided that the lowest fermion mass in the model $m=0$, a condition previously found to be necessary for consistency by Thorn \Tp\ and Schwarz \SchRep. A little later, Ed, Russell Smith (a pure mathematician at Durham), David Olive and I \CGOS\ were able to give an analytic proof of \Deltaexp\ and show that the spectra of states in the dual bosonic channels were  the same apart from reversals of parity. These reversals of parity, implying a doubling of states were unwelcome and, in fact, prior notice that the massless fermions should be chiral, indeed Majorana, as eventually realized by Gliozzi, Scherk and Olive \GSO.

At the same time, Stanley Mandelstam had extended the interacting string picture to the Neveu-Schwarz-Ramond model \Mb\ and obtained the same results on fermion scattering, including the reversals of parity, without having to face the calculation of $M(x)$. Then, and for some years to come, until the work of Polyakov \AP, Mandelstam was virtually alone in using the string picture and functional methods as a basis for calculating amplitudes (but see \RT). 

Then it seemed that the challenges were to find a model for which the critical dimension would be $4$, to understand better amplitudes involving fermions, and to calculate higher order loop contributions. The NSR model, as initially formulated, had a tachyon and a critical dimension $D=10$. It was known that the tachyon could be removed by projecting on to a sector of the model but in a way that made the lowest fermion states, which were massless, chiral. The spectrum of the model necessarily had a massless spin one particle and, in the closed string sector, a massless spin two particle. Thus, although constructed with a view to describing the strong interactions, it had the characteristic massless particles, spin one, spin two and chiral spin one half, of the other interactions: electromagnetism, gravity and the weak interactions, as Ed Corrigan emphasized to me. Ed, interpreting the Ramond fermion as a neutrino in an attempt to sort out the behaviour of the parity of the bosonic states under crossing, even made a calculation of the Weinberg angle, which he talked about at Aspen in August 1974, but did not publish. (See David Olive's review at the London Conference  \Olive\ for some comments on the interpretation of dual models in July 1974.)
  
By the end of 1973, as the fascination of dual models or string theory remained undimmed, though with ever increasing technical demands,  the interest of many was shifting elsewhere. On 21 December, David Olive wrote to me, ``Very few people are now interested in dual theories here in CERN. Amati and Fubini independently made statements to the effect that dual theory is now the most exciting theory that they have seen but that it is too difficult for them to work with. The main excitement [is] the renormalization group and asymptotic freedom, which are indeed interesting." 

In the summer of 1974 I left Durham. At the beginning of 1973, Roger Dashen had invited me to spend one or two years at the Institute for Advanced Study, where Andr\'e Neveu was then a long-term Member. I had obtained leave from Durham for the academic year 1974-75 and then, in April 1974, I was offered and accepted a University Assistant Lectureship in Cambridge, and I was not allowed to delay taking it up beyond January 1975. (Today, I think, a longer delay would be countenanced and perhaps encouraged; research performed would count for the UK Research Assessment Exercise without salary being paid.) 

So, in July 1974, Helen and I, with our nine-month old daughter, Linda,  set out for Aspen, where John Schwarz had assembled a good proportion of the world's dual model/string theorists. We were {\it en route} for Princeton via Berkeley, where I had secured an invitation from Stanley Mandelstam to spend six weeks before the Institute term began. In Berkeley,  I wrote a largely cathartic paper \Gb\ on supersymmetry, which probably helped no one's understanding, except marginally my own. It had one memorable effect: namely, that when I reached Princeton I was invited to give a general seminar on supersymmetry, which most people did not know much about then. When I said I would rather talk about string theory, my offer was politely declined on the grounds that no one in Princeton was interested, a situation that has changed in the intervening years. Somewhat put out by this response, I did not give a seminar at all. 

It was not quite true that no one was interested in strings. Tullio Regge, then a Professor at the Institute, was trying with Andy Hanson and Giorgio Ponzano (whom I never met -- he was back in Torino)  to quantize the relativistic string in a time-like rather than a light-like gauge. We succeeded \GHP\ in doing this by introducing the DDF operators at the classical level and using the formalism of Dirac brackets. Working  with Regge was exhilarating, even if he managed to avoid getting his name on the paper we wrote. 

According to the report I submitted when I left the Institute, much of the remainder of my all too brief stay there (I returned in 1988 for part of a year on conformal field theory but I got a third stay from 2004 only at the cost of becoming Director) was spent trying to to quantize the motion of strings on spaces that were not flat, specifically group manifolds, as a way of compactifying the extra dimensions to produce nonabelian symmetries.  At the time, it seemed that compactification of the unwanted extra dimensions to form a torus (in the absence of a model with $D=4$) would only lead to abelian symmetries. It was almost nine years later before David Olive and I began to see how compactifying on a torus could lead to nonabelian symmetries in string theory, through the affine Kac-Moody algebras formed by vertex operators, and the preferred status of $E_8\times E_8$ and $SO(32)$ \GOa. 

Back in Cambridge in January 1975, I worked with my new research student, Roger Horsley, to try to develop more powerful techniques for handling calculations in dual theory. However, not long after my next student, Nick Manton, began research, I started to realize that following my interests in strings or dual models might be a fine indulgence for me, but it was not going to help my students get jobs.  (One of the great attractions of Cambridge at the time was that chances for promotion were so slim -- Jeffrey Goldstone was still a Lecturer --  that one did not need to be distracted by the prospects for  advancement: they seemed negligible.) 

I visited CERN in the summer of 1976 and there began learning from Jean Nuyts and David Olive and others about magnetic monopoles in gauge theory. (I had attended Gerhard 't Hooft's seminal talk at the London conference in 1974 but I had not understood any of it at the time.) The inverse relation between electric and magnetic charge, first found for electromagnetism by Dirac, generalized to the nonabelian case, but we found that, in general, the magnetic charges were associated with a group somewhat different to the original `electric' gauge group. This led to the introduction of a dual group for the magnetic charges \GNO.
When I met Michael Atiyah for the first time at a conference on mathematical education at Nottingham in the spring of 1977, and told him about results on monopoles, he identified this dual group  with  the dual group that Robert Langlands had introduced for seemingly other reasons some years previously. Only recently are the connections between these occurrences of the dual group being fully understood \AW.

Even though I had felt I should move on from string theory to other topics, for my students' ultimate survival rather than any lack of my own interest, it turned out, perhaps ironically, perhaps inevitably, that the ideas David Olive and I worked in the following ten years ended up being connected with string theory,  even when this was not evident to begin with. Thus, I never escaped, even though I have been distanced by administration for much of the time in the last couple of decades. Many developments over the past forty years have definitively established the credentials of string theory as a deep and significant mathematical structure, while its ultimate status as a physical theory remains to be determined, of course. It is disconcerting to think that this denouement might be reached when one is no longer able to appreciate it. 

\vskip24pt
\noindent{\bf Acknowledgements:}

I am grateful to 
Edward Corrigan, Louise Dolan, David Olive, Siobhan Roberts and Charles Thorn for helpful comments and corrections.

\listrefs

\end